\documentclass[aps,10pt,prl,twocolumn,amsmath,amssymb,floatfix]{revtex4-1}

\usepackage{graphicx}
\usepackage{dcolumn}
\usepackage{bm}
\usepackage{color}
\usepackage[hidelinks,breaklinks]{hyperref}
\usepackage{xcolor}
\hypersetup{
    colorlinks,
    linkcolor={red!50!black},
    citecolor={blue!50!black},
    urlcolor={blue!80!black}
} 
\usepackage{float}

\newcommand{\topic}[1]{\addcontentsline{toc}{subsubsection}{#1}}
\renewcommand{\topic}[1]{}
\usepackage{amsmath}
\usepackage{amssymb}
\usepackage{makeidx}
\usepackage{graphicx}
\usepackage{color}

\newcommand{\hide}[1]{}

\newcommand{\myvec}[1]{\vec{#1}}
\renewcommand{\vr}{{\myvec{r}}}

\renewcommand{\r}{\rangle}
\renewcommand{\l}{\langle}

\renewcommand{\th}{\theta}

\newcommand{\lcase}{\left\{\begin{array}{ll}}
\newcommand{\rcase}{\end{array}\right.}
\renewcommand{\bar}{\begin{array}{ll}}
\newcommand{\ear}{\end{array}}
\newcommand{\bma}{\begin{pmatrix}}
\newcommand{\ema}{\end{pmatrix}}
\newcommand{\beq}{\begin{equation}}
\newcommand{\eeq}{\end{equation}}
\newcommand{\bel}[1]{\begin{equation}\label{eq:#1}}
\newcommand{\eel}{\end{equation}}
\newcommand{\bea}{\begin{eqnarray}}
\newcommand{\eea}{\end{eqnarray}}
\newcommand{\beaNN}{\begin{eqnarray*}}
\newcommand{\eeaNN}{\end{eqnarray*}}

\newcommand{\comm}[1]{{\bf C: #1}}

\newcounter{lecture}

\renewcommand{\comm}[1]{}
\renewcommand{\check}[1]{{\color{green} #1}}

\begin{document}
\title{Dynamic exchange in the strong field ionization of molecules}
\author{Vinay Pramod Majety}
\email[]{vinay.majety@physik.uni-muenchen.de} \author{Armin Scrinzi}
\email[]{armin.scrinzi@lmu.de} \affiliation{Physics Department, Ludwig
  Maximilians Universit\"at, D-80333 Munich, Germany}
 
\begin{abstract}
  We show that dynamic exchange is a dominant effect in strong field
  ionization of molecules. In $CO_2$ it fixes the peak ionization
  yield at the experimentally observed angle of $45^\circ$ between 
  polarization direction and the molecular
  axis. In $N_2$ it changes the alignment dependence of yields by up to a
  factor of 2. The effect appears on the Hartree-Fock level as well as in
  full {\it ab initio} solutions of the Schr\"odinger equation.
\end{abstract}
\maketitle

\topic{Motivation by re-scattering}
  
Experimental techniques like molecular orbital tomography
\cite{Itatani2004,Taeb2012}, laser-driven electron diffraction
\cite{Spanner2004,Xu2014}, and high harmonic imaging
\cite{Smirnova2009} are based on the control of ionization by the strong field
of a laser. They share the concept that an
electron is emitted by a strong laser field and re-directed by the same field to its 
parent system, where it produces a snapshot of the system's time evolution
in the angle-resolved electron-momentum or harmonic spectra. The 
analysis of these experiments relies on the idea that the  
steps of initial electron emission, propagation, and scattering 
of the returning electron can be considered as largely independent.
Adequate understanding of each of these three steps is a
pre-requisite for proper use of the techniques.

In this letter we deal with the ionization step. With atoms, there are several
models that deliver correct ionization yields at infrared
(IR) wavelength. In contrast, for molecules a disquieting discrepancy
between theoretical predictions and experiment appeared: two
independent experiments at two different intensities \cite{PhysRevLett.98.243001,Thomann2008}
reported maximal ionization of $CO_2$ when the molecular axis was
aligned at $45^\circ$ to the polarization direction of a linearly
polarized pulse. In contrast, most theoretical calculations found
angles in the range $30^\circ\sim 40^\circ$.

It is usually assumed that ionization at IR wavelength
is a  tunneling process and yields can be obtained as the integral 
over the tunneling rates computed at the instantaneous field strengths.
As the field ionization rates drop exponentially with the ionization 
potential, one expects that the highest occupied molecular orbital
(HOMO) in a molecule determines ionization. In particular, the angle
dependence of the ionization rate should reflect the electron density distribution 
of the HOMO. Combining this idea with the Ammosov-Delone-Krainov (ADK) \cite{Ammosov1986} formula
for tunneling from effective single-electron systems, the molecular ADK (MO-ADK) approach was
formulated \cite{PhysRevA.66.033402}. In more complicated molecular systems
with energetically closely spaced ionic states this approach
may become invalid \cite{Smirnova2009,Ferre2015}: at
the nodal directions of the HOMO, where MO-ADK would show nearly no ionization, 
the energetically next lower orbital HOMO-1 could contribute.

On this level of theory there remains a striking discrepancy to experimental findings
for the ionization of $CO_2$: while maximum ionization is predicted
for an alignment of molecules at $\sim30^\circ$ between laser polarization and
molecular axis, experiments find the maximum at 45$^\circ$  \cite{PhysRevLett.98.243001,Thomann2008}. 

\topic{DFT, Madsen, Pachkovskii}
Several attempts were undertaken to resolve the discrepancy by more elaborate
computations.
From density functional theory (DFT) calculations it was concluded that
the contribution of energetically lower molecular orbitals cannot account for the experimental 
observation \cite{PhysRevLett.104.223001}. In a time-dependent DFT calculation
reported in Ref.~\cite{PhysRevA.80.011403} peak yield was found at 40$^\circ$ alignment. On the other hand, a 
single electron model with a frozen core potential 
produced the experimental value of  45$^\circ$  \cite{PhysRevA.80.023401}.
 The only fully numerical calculations beyond single
electron or electron density based methods was reported in 
Ref.~\cite{PhysRevA.80.063411} where it was shown that a single channel
picture leads to peak angles  $\sim30^\circ$ and it was conjectured that
inter-channel couplings could explain the experimental observation.
Other efforts using the semi-classical WKB approximation
\cite{PhysRevLett.106.173001} and the strong field eikonal Volkov
approximation \cite{Smirnova2009} also fail to yield accurate
predictions.  A recent work \cite{PhysRevA.91.043406} analyzes the problem
using an adiabatic strong field approximation to show that field-distortion of the
orbitals plays a role, but the predicted angles of $36^\circ$ to $39^\circ$ fall 
short of the experimental values.  In spite of all
efforts, the discrepancy remained unresolved.

\topic{Neglect of exchange}

In the discussion so far, little attention has been paid to 
exchange symmetry that must be respected not only in the initial state but 
also during the ionization process. Ideally, in DFT such effects would be included,
but in practice this is hardly ever fully achieved due to limitations of the 
exchange-correlation potentials. The value reported closest to experiment
was obtained with a single electron potential including DFT-based exchange, 
but the good agreement there was attributed to excited state dynamics rather than exchange \cite{PhysRevA.80.023401}.

\topic{Main contents of the paper: exchange matters, already on HF}

In this letter, we show that exchange occupies a central place 
in strong field ionization (SFI). Specifically, in $CO_2$ the non-local 
exchange forces lead to peak ionization at an
alignment of  45$^\circ$. Effects on the alignment-dependence of $N_2$ ionization 
are sizable but less conspicuous. The mechanism is truly dynamical and independent
of exchange and correlation in the initial states. Qualitatively 
it appears also on the Hartree-Fock level.

We compute SFI rates and solutions of the time-dependent Schr\"odinger equation (TDSE)  
by the {\it ab initio} hybrid
anti-symmetrized coupled channels (haCC) approach \cite{haCC}.
haCC uses a  multi-electron wavefunction in terms
of several ionic states $| I \rangle$ that are 
fully anti-symmetrized with a numerical single electron basis, $|i\rangle $. 
In addition, the neutral ground and excited states,
$|\mathcal{N}\rangle$, can be included resulting in the wavefunction 
\begin{equation} \label{eq:basis} | \Psi_A \rangle = \sum_{i,I}
  \mathcal{A} [ | i \rangle | I \rangle ] C_{i,I} + \sum_{\mathcal{N}}
  | \mathcal{N} \rangle C_{\mathcal{N}},
\end{equation}
which we will refer to as ansatz A in the following.
The $C_{i,I}, C_{\mathcal{N}}$ are the respective expansion 
coefficients and $\mathcal{A}$ indicates anti-symmetrization.  
The $|I\r$ and $|\mathcal{N}\r$ states were obtained from the 
COLUMBUS quantum chemistry package \cite{Lischka2011}. 
For $|i\r$ we use a high-order finite element radial basis
combined with single center spherical harmonics. 
A complete description of the method can be found in
\cite{haCC}. It is important that the basis can accurately 
describe the asymptotic behavior of the ionizing orbital, as discussed in Ref.~\cite{PhysRevA.80.051402}.
Neutral and ionic states can be systematically included to examine multi-electron effects like field-free 
correlation, inter-channel coupling and ionic core polarization. Independently the importance of exchange can also be 
investigated.

Tunneling ionization rates are computed using  exterior complex scaling
\cite{simon79:complex-scaling, Reinhardt82, PhysRevLett.83.706}: the Hamiltonian
is analytically continued by transforming the electron coordinates into the complex plane. 
For radii $r>R_0$ one uses $r_\theta = e^{i\theta}(r-R_0) + R_0$ with the complex
scaling angle $\theta>0$. The resulting Hamiltonian is non-Hermitian 
with a complex ground state eigenvalue $W=E_0+E_s-\frac{i}{2}\Gamma$,
where $E_0$ is the field-free ground state energy, $E_s$ is its dc-Stark shift and
$\Gamma/\hbar$ is the static field ionization rate.
Apart from errors due to finite computational approximation, $W$ is independent of $\th>0$ and $R_0\geq 0$. 
     
We treat the $CO_2$ molecule with nuclear positions fixed at the equilibrium {\it C-O} bond length of
116.3 pm. The multi-electron states of neutral and ion are computed using COLUMBUS with the minimally augmented cc-pvtz basis 
at the multi-reference configuration interaction singles level. We used up to 6 ionic channels 
which include the doubly degenerate $X^2\Pi_g$,
$A^2\Pi_u$, and the singly degenerate $B^2\Sigma_u^+$, $C^2\Sigma_g^+$
states. Single electron functions with up to 84 linear coefficients with finite element orders 12 
on a radial box of 30 a.u and up to 269 spherical harmonics ($L_{max} =
12,M_{max}=12$) were used for the stationary problem. For solutions of the TDSE the 
number of spherical harmonics was increased up to 324. This numerical basis is complemented by the atom-centered 
Gaussians that constitute the neutral and ionic functions.
For complex scaling, we chose $R_0$-values well outside the range of neutral and ionic orbitals,
such that only the coordinate of the single-electron basis had to be continued to complex values. 
Basis and scaling the parameters $R_0$ and $\th$  were systematically varied 
to ensure that results are converged 
to better than 2\%. The main approximation is introduced by the limited number of 
ionic channels. With all 6 ionic channels, we obtain a first ionization potential of 
$I_p=13.85eV$ (Experimental value: 13.78eV \cite{Ehara1999}), which decreases
by about 0.14 eV with fewer ionic channels. 
  
\topic{Main results: agrees with experiment, profiles as function of intensity, multi-electron}

\begin{figure}[htb]
     \centering
     \includegraphics[scale=0.52]{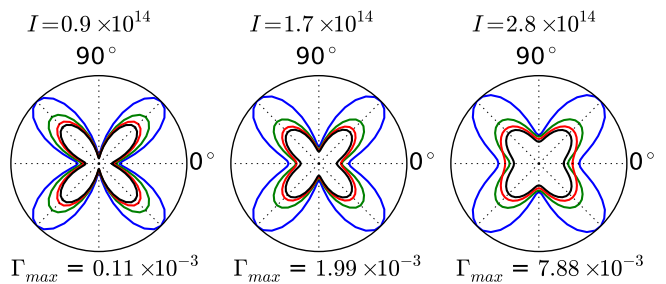}
     \caption{Alignment angle dependent $CO_2$ ionization rates at selected
       intensities $I$ (in $W/cm^2)$. The convergence with the number of ionic channels indicates
the role of multi-electron effects. Blue: including only the neutral ground state and ionic $X^2\Pi_g$ ground states,
       green: as blue with the ionic $A^2\Pi_u$ channel added.  Red: as green with $B^2\Sigma_u^+$ channel. Black:
        as red with $C^2\Sigma_g^+$ channel. Computations were performed for static fields of strengths
       $F=0.05$, 0.07 and 0.09 atomic units corresponding to intensities $I=F^2/2$ that label the plots.  
       $\Gamma_{max}$ indicates maximal decay width in atomic units at the inclosing circle. A total of 6 ionic channels are used in the
       calculations.
}
     \label{fig:co2_full}
   \end{figure}

\begin{figure}[htb]
      \centering
      \includegraphics[width=0.45\textwidth]{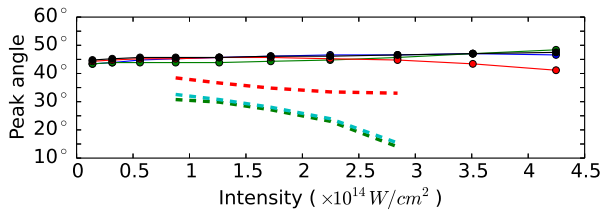}
       \caption{Peak ionization angles as a function of intensity. Solid lines: results with the 
anti-symmetrized ansatz A, Eq.~(\ref{eq:basis}). 
Dashed lines: results without anti-symmetrization, ansatz B, Eq.~(\ref{eq:basis_1}). 
Colors correspond to different numbers of neutral states and ionic channels, see Fig.~\ref{fig:co2_full} (solid) and Fig.~\ref{fig:co_1} (dashed).
}
      \label{fig:co2_max_ang}
 \end{figure}

The main results of this letter are given in figures
\ref{fig:co2_full} and \ref{fig:co2_max_ang}.
In Fig.~\ref{fig:co2_full},  one sees that the static field ionization rates
peak at an alignment angle of 45$^{\circ}$. They have
minima at 0$^{\circ}$ and 90$^{\circ}$ due to the nodal planes in the
HOMO orbital of $CO_2$. These findings are in good agreement with
experiments \cite{PhysRevLett.98.243001,Thomann2008}.
As the dominant multi-electron effect, we see a reduction of ionization rates
as the number of ionic channels is increased. This can be understood as
a larger number of ionic channels results in larger polarizability
of the ground state, whereas the more tightly bound ionic states are less affected. 
The difference in dc-Stark shifts increases the ionization
potentials in presence of the field  and the ionization rates drop. 
 
Figure \ref{fig:co2_max_ang} shows the angle of peak rate as a function 
of intensity: except for the highest intensities, the rate varies by $\lesssim2^{\circ}$, 
depending on the number of ionic 
channels included. We cannot confirm any intensity dependence
as was predicted in Ref.~\cite{PhysRevLett.106.173001} based on analytic arguments.  
Dependence on the number of channels is strongest at the higher intensities $I\gtrsim 2.5 \times 10^{14} W/cm^2$.
There, the tunneling picture ceases to be applicable:  
according to a simple estimate \cite{PhysRevLett.83.706} at intensities  
$I_{\rm b}\approx I_p^2/4 = 1.5\times 10^{14} W/cm^2$ the molecular binding barrier of $CO_2$ 
is suppressed to below the field free ground state energy. 
In this regime, the importance of virtual continuum states for polarization of the ionic core may become important, 
which is not modeled by the haCC ansatz as used here and therefore no dependable statement
about the accuracy of our results can be made.  
   
\topic{Fig: rates and channel dependence}
  
\topic{Valitidy of QSA}

\begin{figure}[htb]
      \includegraphics[scale=0.35]{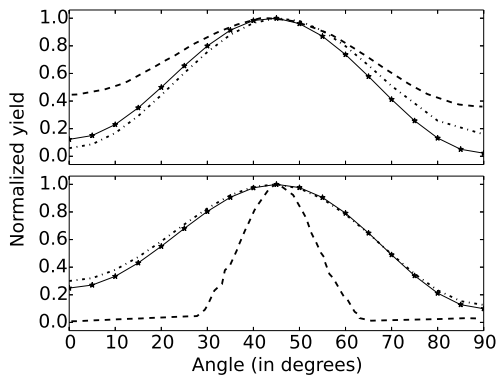}
      \caption{Normalized angle dependent yields from TDSE (lines),
        QSA (dash-dotted lines) in the single channel picture and
        experiments \cite{PhysRevLett.98.243001,Thomann2008} (dashed
        lines). The laser parameters are 800 nm central wavelength, 40
        fs duration with peak intensities of 3 $\times 10^{13} W/cm^2$
        (Upper panel) and 1.1 $\times 10^{14} W/cm^2$ (lower
        panel).}
      \label{fig:quasi_stat}
    \end{figure}

The alignment dependence of ionization obtained in quasi-static approximation (QSA) 
by integrating the tunnel ionization rate is confirmed by solutions of the complete TDSE.  
In figure \ref{fig:quasi_stat}, normalized angle dependent yields obtained from TDSE and QSA
    within the single channel model are compared with experimental
    results. The angle-dependence in TDSE is well approximated in QSA,
with better agreement for higher intensities, where the QSA is more appropriate \cite{PhysRevLett.83.706}.
This agreement is gratifying, considering that in the intensity range
$3\times10^{13}-1.1\times10^{14}W/cm^2$ with Keldysh parameters $\gamma=2\sim1$,
one can hardly expect ionization to be of pure tunneling type.
A failure of the tunneling picture is exposed in the {\em magnitudes} 
of the yields, where the TDSE results exceeds the QSA by a factor 2 at $1.1\times10^{14}W/cm^2$
and by nearly two orders of magnitude at $3\times10^{13}W/cm^2$.
The peak angle is consistent with the experiments, but yields
found  in one of the experiments~\cite{PhysRevLett.98.243001} are more narrowly confined
around the maximum angle. It was noted in Ref.~\cite{PhysRevA.80.051402} that
the experimental result may be artificially narrowed due to the deconvolution procedure.
   
\topic{Fig: TDSE vs. QSA vs. Experiment }
   
\topic{commenting on failure in literator}

\begin{figure}[htb]
    \centering
    \includegraphics[scale=0.52]{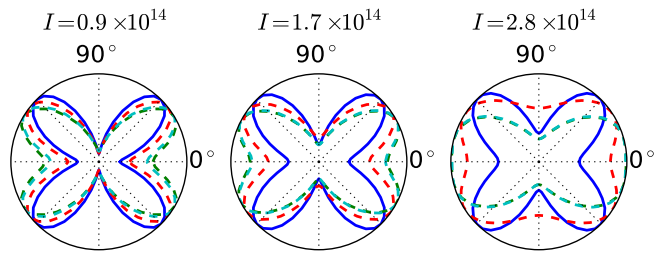}
    \caption{The role of  exchange in $CO_2$ ionization: alignment angle-dependence of normalized static SFI rates in different single-channel models. 
 Blue: anti-symmetrized ansatz A with the neutral ground state and ionic $X^2\Pi_g$ ground state channels. 
       Green: ansatz B with the same states as blue, red: as green, with the addition of ionic $A^2\Pi_u$ state. 
      Cyan: as green with the addition of first excited neutral state. The green and cyan lines coincide at the two higher intensities.
      }
    \label{fig:co_1}
  \end{figure}

The failure of earlier theory in reproducing the peak angle of $45^\circ$
is due to the absence or insufficient inclusion of dynamical exchange. 
This is clearly seen by omitting from the haCC ansatz A the 
anti-symmetrization of the single-electron basis against the multi-electron states 
in an otherwise identical wavefunction, ansatz B:
    \begin{equation} \label{eq:basis_1} | \Psi_B \rangle = \sum_{i,I} |
      i \rangle | I \rangle C_{i,I} + \sum_{\mathcal{N}} | \mathcal{N}
      \rangle C_{\mathcal{N}},
    \end{equation}
In figure \ref{fig:co_1} one sees that with ansatz B one obtains the peak rate at an angle
around 30$^{\circ}$ at low intensities that then dips-off as intensity is increased, 
see also Fig.~\ref{fig:co2_max_ang}.

Our results without anti-symmetrization for the dynamics 
are consistent with Ref.~\cite{PhysRevA.80.063411}, where it was proposed that
the remaining discrepancy to the experimental value was caused by neglecting
coupling between $X^2\Pi_g$, $A^2\Pi_u$ ionic channels in the 
calculation. In contrast, in Ref.~\cite{PhysRevA.80.023401},
the angle near $45^\circ$ was attributed to dynamics of excited neutral states, mostly 
the first excited neutral state. 
However, neither excited state dynamics nor 
coupling of ionic channels, in absence of dynamical exchange, result in correct angles.  

Fig.~ \ref{fig:co_1} shows that the first excited state of the neutral has hardly 
any discernable role in determining the emission profile and does not influence the angle 
of peak emission.
Coupling of  channels as proposed in \cite{PhysRevA.80.063411} does move the angle closer 
to experiment, but still does not yield the correct result. The improvement can be understood as, 
in the limit of a complete set of channels, ansatz A and B are equivalent. However, the primary 
role of the seemingly complicated multi-electron dynamics is to mimic dynamical exchange. In contrast, 
with dynamical exchange properly considered,  
a simple essentially single-electron picture of field-ionization re-emerges. 

\begin{figure}
   \includegraphics[height=0.17\textwidth]{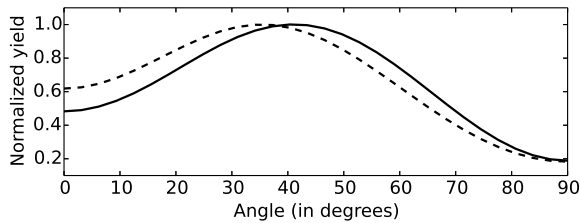}
   \caption{Rate as a function of alignment, computed with Hartree-Fock 
neutral ground state, ionic ground state in Koopman's approximation, and Coulomb potential truncated at 10~a.u. 
Solid, with exchange,  ansatz A, and dashed, without dynamical
exchange, ansatz B. Field strength = 0.06 a.u.}
   \label{fig:hf}
 \end{figure}

We demonstrate this by reducing the problem to the simplest possible case. 
We use the Hartree-Fock neutral state of $CO_2$ and the ion ground state in Koopman's approximation.
Denoting by $\{\phi_1,..,\phi_N \}$ the occupied Hartree-Fock
orbitals of the neutral and by $\psi(t)$ the active electron, ansatz A and B are reduced
to
\bea
|\Psi_A\r&=&\det(|\psi(t)\r|\phi_2\r\ldots|\phi_N\r)C_{11}+|\mathcal{N}\r C_\mathcal{N}\\
|\Psi_B\r&=&|\psi(t)\r\det(|\phi_2\r\ldots|\phi_N\r)C_{11}+|\mathcal{N}\r C_\mathcal{N},
\eea
where $\det$ indicates the Slater determinant.
The effective Hamiltonians governing the time-evolution of $\psi(t)$ for the two cases 
differ only by the exchange term
 \begin{equation}
     (V_{\rm x}\psi)(\vr)=
     \sum_{k=2}^N \phi_k(\myvec{r}) \int\,d^3r' \frac{\phi_k(\myvec{r}')\psi(\myvec{r}')}{|\myvec{r}-\myvec{r}'|}.
 \end{equation}
In Ref.~\cite{PhysRevA.80.051402} it was pointed out that the long-range interactions also affect emission.
To exclude those, we smoothly truncate the Coulomb tail of the 
potential at 10~a.u.\  Figure \ref{fig:hf} shows that also here exchange shifts the peak angle by $\sim7^\circ$.

Apart from the exchange term, ansatz A effectively enforces orthogonality of the active electron orbital against the ionic
HF orbitals $\l \psi|\phi_k\r = 0, k \geq 2$. If this were the dominant effect of anti-symmetrization, one would expect
that in absence of the constraint (ansatz B) the ground state energy would be lowered. On the other hand,
anti-symmetrization effectively enlarges the ansatz space: it operates in the $N$-fold larger space containing 
all permutations of $\psi$ through the $\phi_2\ldots\phi_N$, but including explicitly only the dynamically accessible 
subspace of anti-symmetrized linear combinations. By this reasoning, Stark-shift (polarization) should be larger 
in ansatz A. Indeed, we find the latter in our calculations. We also directly verified that an orthogonality constraint
on $\psi(t)$ against the $\phi_k$ in ansatz B causes only $\lesssim1\%$ of the overall difference
between the results of A and B. This finally establishes that indeed the dynamical effects of exchange
play the decisive role in ionization. 
  
\begin{figure}[htb]
     \centering
     \includegraphics[scale=0.52]{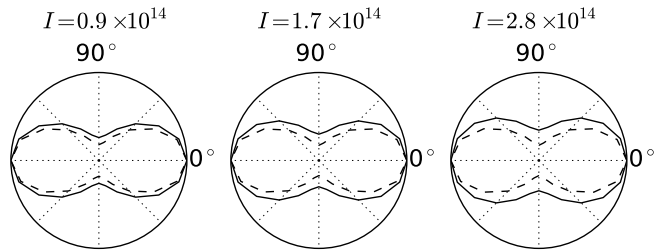}
     \caption{Normalized ionization rates of
       $N_2$ as a function of alignment angle. 
Including neutral ground state and the $X^2\Sigma_g^+$ ionic ground state channel. Solid: with dynamic exchange, ansatz A,
and dashed, without exchange, ansatz B.}
     \label{fig:N2}
   \end{figure}

Dynamical exchange is most conspicuous in $CO_2$ because of the 
node-structure of the HOMO and the resulting non-trivial angle dependence
of the yields. However, the mechanism as such is universal and must be included
for obtaining correct ionization rates from any system. As an example, we studied the
effect on the nitrogen molecule, which is one of the main model system for strong field 
physics. Figure \ref{fig:N2}
 shows normalized ionization rates for $N_2$ at equilibrium nuclear position 
 with a single channel in ansatz A and B. 
Here, dynamical exchange leads to a broadening of the ionization profile,
where the ratio between the rates at $0^\circ$ and $90^\circ$ changes by up to a factor $\sim 2$.  
  
In conclusion, we have established that dynamical exchange takes a central place in the
ionization of molecules. The effects on $CO_2$ are striking, but also for $N_2$ results can change by 
up to a factor 2 merely due to exchange. This indicates that dynamical exchange must be considered 
in any attempt to understand strong field ionization also of more complex multi-electron systems.
Apart from the ionization yields discussed here, the angular distribution of
electron emission at fixed alignment may be affected. A critical assessment of the importance of these
distributions for rescattering-based attosecond experiments appears in place. On the other hand,
simple anti-symmetrization may enhance single-electron and single-channel models that
have been applied so far, even without the comparatively heavy numerical apparatus used to establish
the fact in the present paper.
    
   The authors acknowledge financial support from the EU Marie Curie
   ITN CORINF, German excellence cluster - Munich Advanced
   Photonics, and by the Austrian Research Fund (ViCoM, F41).

   \bibliography{co2_paper}{} \bibliographystyle{unsrt}
   \bibliographystyle{apsrev4-1}
  
 \end{document}